# Walk model that continuously generates Brownian walks to Lévy walks depending on destination attractiveness

Shuji Shinohara[a,*], Daiki Morita[a], Hayato Hirai[a], Ryosuke Kuribayashi[a], Nobuhito Manome[b,c], Toru Moriyama[d], Hiroshi Okamoto[b], Yoshihiro Nakajima[e], Yukio-Pegio Gunji[f], and Ung-il Chung[b]

[a] School of Science and Engineering, Tokyo Denki University, Saitama, Japan

[b] Department of Bioengineering, Graduate School of Engineering, The University of Tokyo, Tokyo, Japan

[c] Department of Research and Development, SoftBank Robotics Group Corp., Tokyo, Japan

[d] Faculty of Textile Science, Shinshu University, Ueda, Japan

[e] Graduate School of Economics, Osaka City University, Osaka, Japan

[f] Department of Intermedia Art and Science, School of Fundamental Science and Technology, Waseda University, Tokyo, Japan

* Corresponding author

E-mail: s.shinohara@mail.dendai.ac.jp

Postal address: School of Science and Engineering, Tokyo Denki University, Ishizaka, Hatoyama-machi, Hiki-gun, Saitama 350-0394, Japan

*Abstract*

The Lévy walk, a type of random walk characterized by linear step lengths that follow a power-law distribution, is observed in the migratory behaviors of various organisms, ranging from bacteria to humans. Notably, Lévy walks with power exponents close to two, also known as Cauchy walks, are frequently observed, though their underlying causes remain elusive.

This study proposes a walk model in which agents move toward a destination in multi-dimensional space and their movement strategy is parameterized by the extent to which they pursue the shortest path to the destination. This parameter is taken to represent the attractiveness of the destination to the agents.

Our findings reveal that if the destination is very attractive, agents intensively search the area around it using Brownian walks, whereas if the destination is unattractive, they explore a distant region away from the point using Lévy walks with power exponents less than two. In the case where agents are unable to determine whether the destination is attractive or unattractive, Cauchy walks emerge. The Cauchy walker searches the region with a probability inversely proportional to the distance from the destination. This suggests that it preferentially searches the area close to the destination, while concurrently having the potential to extend the search area much further.

Our model, which can change the search method and search area depending on the attractiveness of the destination, has the potential to be utilized for exploring the parameter space of optimization problems.

# INTRODUCTION

Lévy walks have been observed in the migratory behaviors of organisms across a range of scales, from bacteria and T cells to humans.[1–5] These walks, a specialized type of random walk, exhibit step lengths $l$ that follow a power-law distribution $P(l) = al^{-\mu}, 1 < \mu \leq 3$, in contrast to the exponentially distributed step lengths of the Brownian walk (where the frequency of step length $l$ is characterized by an exponential distribution $P(l) = \lambda e^{-\lambda l}$). Lévy walks are particularly noted for their occasional, very long linear movements. Frequently, Lévy walks with exponents close to two have been documented in various organisms, sparking interest in the reasons behind these patterns.[1,7–12] Such walks, when the exponent is two, are also known as Cauchy walks. The Lévy flight foraging hypothesis (LFFH)[13,14] suggests that under conditions where food is scarce and randomly dispersed, and predators lack any memory of food locations, Cauchy walks represent the most efficient foraging strategy and offer evolutionary benefits.[15] Historically, it has been accepted that search efficiency peaks for inverse square Lévy walks with an exponent of two as per the LFFH.[16] However, recent studies challenge this assumption, showing that Cauchy walks only achieve maximum search efficiency under specific conditions in multi-dimensional spaces.[17] Conversely, research by Guinard and Korman has demonstrated that an intermittent Cauchy walk becomes the optimal search strategy in finite two-dimensional domains, particularly when the objective is to quickly locate targets of any size.[18] The ongoing debates continue to explore the natural conditions and search methodologies that render the Cauchy walk optimal.

Although Lévy walks have traditionally been linked to the execution of an optimal search strategy for sparsely and randomly distributed resources, this interpretation has not been universally accepted.[19–21] Offering a different perspective from the LFFH, Abe proposed that the functional advantages of Lévy walks stem from the critical phenomena within the system, demonstrating that Lévy walks emerge near a critical point between stable synchronous and unstable asynchronous states.[22] This occurrence is significant because, near the critical point, the range of inputs from which information can be discriminated is broader, providing organisms the flexibility to alternate between searching for nearby resources and venturing toward new, distant locations based on the received inputs. In Abe's model, while Lévy walks appear near the critical point, they do not conform to a Cauchy walk with an exponent of two. Conversely, Sakiyama developed an algorithm that effectively generates Cauchy walks through the decision-making process of a single walker.[11]

As highlighted in the LFFH, Lévy walks are not universally applicable across all environments or conditions. Humphries et al. found that Lévy behavior is associated with environments where prey is sparse, whereas Brownian movements correlate with areas where prey is abundant.[4] Similarly, de Jager et al. observed that Brownian motion arises from frequent interactions among organisms in densely populated environments; they also noted that in controlled experiments adjusting for population density, the movement patterns of mussels transitioned from Lévy to Brownian motion as density increased.[23] Additionally, Huda et al. demonstrated that metastatic cells exhibit Lévy walks, whereas non-metastatic cancer cells engage in simple diffusive movements.[12] Huo et al. reported that the movements of *Escherichia coli* cells are super diffusive, aligning with Lévy walk behavior. In contrast, they observed that mutant cells lacking chemotaxis signaling noise displayed normal diffusive trajectories.[24] From these observations, they concluded that Lévy walks stem from the noise associated with chemotaxis signaling.

The primary aim of this study is to develop a straightforward, abstract walk model that consistently produces Cauchy walks and to identify the specific conditions under which these walks appear. Consider an agent navigating toward a destination in two-dimensional space. We introduce a function $z$, which depends on the distance from the agent's current position to the destination and increases in value as this distance decreases. By implementing this function, the agent's original goal of approaching the destination is transformed into the objective of maximizing $z$. In other words, $z$ can be said to be a function that expresses the degree of satisfaction for agents who want to get closer to the destination.

The agent aims to shift the position $(x, y)$ to increase $z$ by a small amount $\Delta z$. The task here is to obtain

$\Delta x$ and $\Delta y$ such that $\Delta z = z(x + \Delta x, y + \Delta y) - z(x, y)$. When $\Delta z$ is very small, the amount of movement of $x$ required to realize the objective can be approximated as $\Delta x \approx \Delta z \frac{\partial x}{\partial z}$ using partial differentiation. Similarly, the amount of movement of $y$ can be approximated as $\Delta y \approx \Delta z \frac{\partial y}{\partial z}$. Although only an approximation, increasing $z$ by $\Delta z$ can be realized by moving only one of $x$ or $y$, or by moving both $x$ and $y$ in any allocation. In other words, the purpose of increasing $z$ to $z + \Delta z$ can be achieved in multiple ways and the way cannot be uniquely determined.

We now turn our attention to some strategies that allocate the amount of modification by $\beta$ to $1 - \beta$ ratio in both the $x$- and $y$-directions as $\Delta x \approx \beta \Delta z \frac{\partial x}{\partial z}$ and $\Delta y \approx (1 - \beta) \Delta z \frac{\partial y}{\partial z}$, respectively. Specifically, we consider the behavior of two cases: a strategy that sets $\beta$ randomly, and a strategy that adjusts $\beta$ such that movement $l = \sqrt{(\Delta x)^2 + (\Delta y)^2}$ is minimized similarly as in the steepest descent method. The first strategy can be classified as a non-minimum displacement strategy, as it does not aim to minimize the amount of movement. Conversely, the second strategy is a minimum displacement strategy, focusing on reducing movement as much as possible. We propose a general agent model that continuously connects the two strategies by introducing a certain control parameter $\gamma$. Parameter $\gamma$ can be regarded as a scale representing the attractiveness of the destination for an agent, wherein an agent with the minimum displacement strategy tries to reach the destination by the shortest route, whereas an agent using the non-minimum displacement strategy tries to approach the destination slowly, taking a detour.

Our theoretical analysis indicates that when the non-minimum strategy with $\gamma = 0.0$ is used, the frequency distribution of movement $l$ follows a power-law with an exponent of two, characteristic of a Cauchy walk. In contrast, employing the minimum strategy with $\gamma = 1.0$ results in a Brownian walk.

# METHODS

## Walk model

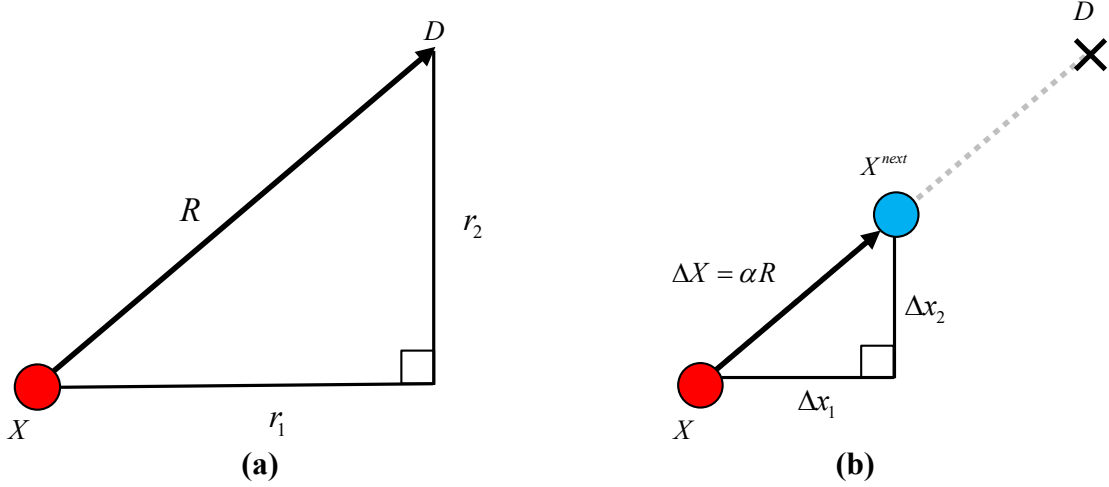

**Figure 1. Diagram of walk model. (a) Agent tries to approach destination *D* from its current position *X*. (b) WA is an algorithm that moves *X* close to *D*.**

This paper deals with an agent walking in a multi-dimensional space. Let us denote the agent's current position vector as $X = \left( x_1, \cdots, x_i, \cdots, x_N \right)^T$ and destination vector as $D = \left( d_1, \cdots, d_i, \cdots, d_N \right)^T$. The difference vector between *D* and *X* is denoted by $R = D - X = \left( d_1 - x_1, \cdots, d_i - x_i, \cdots, d_N - x_N \right)^T = \left( r_1, \cdots, r_i, \cdots, r_N \right)^T$. The norm of the vector is denoted by $r = \| R \| = \sqrt{\sum_{i=1}^{N} \left( r_i \right)^2}$ (Fig. 1(a)).

The position of the agent at the next time is $X_{next}$, and the movement vector is $X_{next} - X = \Delta X = \left( \Delta x_1, \cdots, \Delta x_i, \cdots, \Delta x_N \right)^T$. The step length is defined as $l = \| \Delta X \| = \sqrt{\sum_{i=1}^{N} \left( \Delta x_i \right)^2}$.

The simplest walking model would be to assume that destination *D* is the agent's position $X_{next}$ at the next time, i.e., $X_{next} = D = X + R$. We model an agent that attempts to approach destination *D* rather than moves directly to it. The simplest algorithm for approaching *D* is to use the weighted average (WA) of *X* and $D$. The WA model is defined as follows (Fig. 1(a)).

$$
\begin{aligned}
X_{next} &= (1 - \alpha) X + \alpha D \\
&= X + \alpha (D - X) \qquad (1) \\
&= X + \alpha R
\end{aligned}
$$

Here, $0 \leq \alpha \leq 1$ represents the change rate.

The model that directly reaches $D$ corresponds to the case where $\alpha = 1.0$ in WA. As illustrated in Fig. 1(b), the agent using WA employs the minimum displacement strategy. Throughout this paper, agents utilizing this strategy will be termed Min agents, while those adopting the non-minimum displacement strategy will be referred to as non-Min agents.

To extend the Min agent to the non-Min agent, we introduce a function $z$ which calculates the input as the distance from the agent's current position to the destination, outputting a higher value as this distance decreases.

$$
\begin{aligned}
z(x_1, \cdots, x_i, \cdots, x_N) &= \exp \left( - \frac{\sum_{i=1}^{N} (d_i - x_i)^2}{2\Sigma} \right) \\
&= \exp \left( - \frac{\sum_{i=1}^{N} (r_i)^2}{2\Sigma} \right) \qquad (2) \\
&= \exp \left( - \frac{r^2}{2\Sigma} \right)
\end{aligned}
$$

Here $0 < z \leq 1$ and the height of the top $z(d_1, \cdots, d_i, \cdots, d_N)$ is 1. $\Sigma > 0$ is a parameter that represents the spread of $z$ as well as the variance of the normal distribution. By introducing $z$, the objective of approaching $D$ is replaced by the objective of increasing $z$. In other words, $z$ can be said to be a function that expresses the degree of satisfaction for agents who want to get closer to $D$. In this paper, $\Sigma$ is set so that $r^2 \ll 2\Sigma$ to simplify later analysis. By setting $\Sigma$ in this way, it is possible to approximate that $z = \exp \left( - \frac{r^2}{2\Sigma} \right) \approx 1 - \frac{r^2}{2\Sigma}$ using the Maclaurin expansion.

Define the movement vector $\Delta X$ when the non-Min agent attempts to increase $z$ by $\Delta z$ as follows.

$$\Delta x_i = 2\alpha\beta_i\Delta z\frac{\partial x_i}{\partial z} \qquad (3)$$

$\beta_i$ for the non-Min agent was set randomly, where $\beta_i$ satisfies the conditions $0 \leq \beta_i \leq 1$ and $\sum_{i=1}^{N}\beta_i = 1$.

$\Delta z$ is the difference between the top and current height, defined as

$$\Delta z = z\left(d_1, \cdots, d_i, \cdots, d_N\right) - z\left(x_1, \cdots, x_i, \cdots, x_N\right)$$
$$= 1 - z\left(x_1, \cdots, x_i, \cdots, x_N\right) \qquad (4)$$

When $r^2 \ll 2\Sigma$, we can approximate $z = \exp\left(-\dfrac{r^2}{2\Sigma}\right) \approx 1 - \dfrac{r^2}{2\Sigma} = 1 - \dfrac{\sum_{i=1}^{N}\left(d_i - x_i\right)^2}{2\Sigma}$, so $\dfrac{\partial z}{\partial x_i} \approx \dfrac{r_i}{\Sigma}$,

$\Delta z = 1 - z \approx \dfrac{r^2}{2\Sigma}$. Equation (3) can therefore be approximated as follows:

$$\Delta x_i = 2\alpha\beta_i\Delta z\frac{\partial x_i}{\partial z}$$
$$\approx \alpha\beta_i\frac{r^2}{r_i} \qquad (5)$$

From Eq. (5), since $r_i\Delta x_i \approx \alpha\beta_i r^2$ and $\sum_{i=1}^{N}\beta_i = 1$ are satisfied, $R \cdot \Delta X = \sum_{i=1}^{N}r_i\Delta x_i \approx \sum_{i=1}^{N}\alpha\beta_i r^2 = \alpha r^2$

holds. In particular, $\Delta X$ is approximately a point on the hyperplane

$H\left(R, \alpha r^2\right) = \left\{\Delta X \in \mathbb{R}^N \mid R \cdot \Delta X = \alpha r^2\right\}$ whose normal vector is $R$ (Fig. 2).

The movement vector of the Min agent can be defined from Eq. (1) as follows:

$$\Delta x_i = \alpha r_i. \qquad (6)$$

Comparing Eqs. (5) and (6), $\beta_i$ is expressed by

$$\beta_i = \frac{r_i}{2\Delta z}\frac{\partial z}{\partial x_i}$$
$$\approx \frac{r_i^2}{r^2} \qquad (7)$$

Thus, the Min agent corresponds to the special case where $\beta_i$ is given by Eq. (7) in the non-Min agent.

Specifically, the non-Min agent moves somewhere onto the hyperplane $H$, whereas the Min agent moves to one specific point on the hyperplane $H$; i.e., the intersection of $H$ and $\alpha R$. Note that, as shown in Fig. 2, in some cases the non-Min agent could be far from the destination.

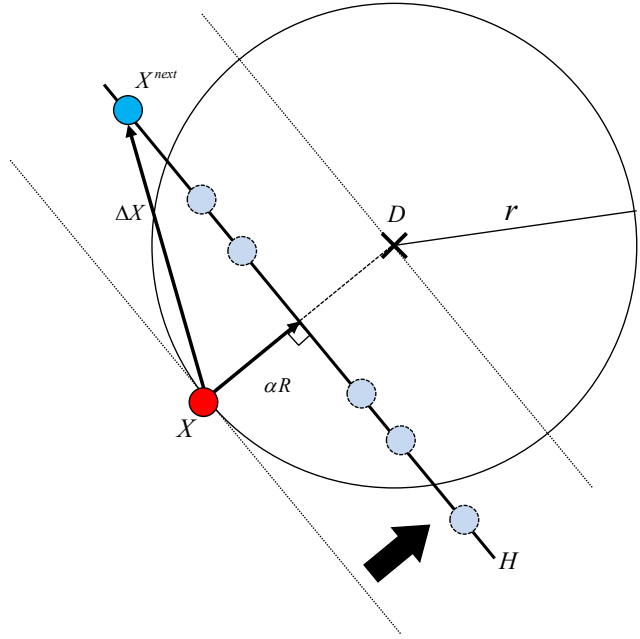

**Figure 2. The non-Min agent moves somewhere onto the hyperplane $H$ at the next time. The Min agent moves to one specific point on the hyperplane $H$; i.e., the intersection of $H$ and $\alpha R$.**

Next, we propose the following general model that continuously connects non-Min agents and Min agents.

$$\Delta x_i = 2\alpha\eta_i\Delta z \frac{\partial x_i}{\partial z}$$
$$\approx \alpha\eta_i \frac{r^2}{r_i} \tag{8}$$

$$\eta_i = \beta_i^{1-|\gamma|}\left(\frac{r_i}{2\Delta z}\frac{\partial z}{\partial x_i}\right)^{\gamma}$$
$$\approx \beta_i^{1-|\gamma|}\left(\frac{r_i^2}{r^2}\right)^{\gamma} \tag{9}$$

Here, $\gamma = 1.0$ corresponds to the Min agent and $\gamma = 0.0$ corresponds to the non-Min agent.

Assume that the walking speed of the agent is finite, and limit the maximum distance traveled per unit time to $l_{max} \geq l = \|\Delta X\|$. Thus, Eq. (8) can be rewritten as follows:

$$\begin{cases} \Delta x_i \approx \alpha \eta_i \dfrac{r^2}{r_i} & if \ l_{max} \geq l \\ \Delta x_i \approx \alpha \eta_i \dfrac{r^2}{r_i} \dfrac{l_{max}}{l} & otherwise \end{cases} . \qquad (10)$$

The agent has the freedom to set the direction of each axis, resulting in each axis being randomly set each time. Let us denote the $N$-dimensional standard basis as $\{e_1, \cdots, e_i, \cdots, e_N\}$. Here $e_i$ represents an $N$-dimensional fundamental vector where the $i$-th element is 1 and all other elements are 0.

Initially, a new $N$-dimensional orthonormal system, $\{e'_1, \cdots, e'_i, \cdots, e'_N\}$, is generated using the Gram-Schmidt orthogonalization method. In this context, a relationship is established between the difference vector $R = (r_1, \cdots, r_i, \cdots, r_N)^T$ in the standard basis and the difference vector $R' = (r'_1, r'_2, \cdots, r'_N)^T$ in the new orthonormal system.

$$(e'_1 \cdots e'_i \cdots e'_N)(r'_1, \cdots, r'_i, \cdots, r'_N)^T = (e_1 \cdots e_i \cdots e_N)(r_1, \cdots, r_i, \cdots, r_N)^T \qquad (11)$$

If the transformation matrix between the two orthogonal systems is $A = (a_{ij})$ and

$(e'_1 \cdots e'_i \cdots e'_N) = (e_1 \cdots e_i \cdots e_N) A$, then $A = (e_1 \cdots e_i \cdots e_N)^{-1} (e'_1 \cdots e'_i \cdots e'_N)$.

For the standard basis, $(e_1 \cdots e_i \cdots e_N)$ and $(e_1 \cdots e_i \cdots e_N)^{-1}$ are the identity matrices, and hence,

$A = (e'_1 \cdots e'_i \cdots e'_N)$. From Eq. (11), it can be observed that the relationship between $AR' = R$ and

$R' = A^{-1} R$ can be established.

The specific procedure for calculating the movement vector is described below. Initially, a new destination vector $D$ and a new $N$-dimensional orthonormal system are generated. Next, using the transformation matrix, the difference vector $R$ in the standard basis is converted to the difference vector $R' = A^{-1} R$ in the new orthonormal system. Subsequently, Eqs. (9) and (10) are employed to derive the movement vector $\Delta X' = (\Delta x'_1, \cdots, \Delta x'_i, \cdots, \Delta x'_N)^T$ in the new orthonormal system. Next, $\Delta X = A \Delta X'$ is utilized to return $\Delta X'$ to the movement vector $\Delta X$ in the standard basis. Finally, the agent's current position is updated to $X + \Delta X$. This process is repeated iteratively.

## Simulation setting

The simulation is performed in a two-dimensional space, and deals with two types of destinations. First, assume that the agent randomly searches its own surroundings. To accomplish this, the direction $\theta$ of the destination is sampled from the uniform distribution of $[0, 2\pi)$, and the distance $r$ from the agent to the destination is sampled from the one-dimensional exponential distribution $P(r) = \lambda e^{-\lambda r}$. Finally, let $D = \left(x_1 + r\cos\theta, x_2 + r\sin\theta\right)^T$. Simulations that use this type of destination are henceforth referred to as Simulation 1.

Second, we fix the true destination at the origin $\left(0, 0\right)^T$. However, we assume that some identification error exists, and $D$ is sampled each time from the normal distribution $N\left(D \,|\, \left(0, 0\right)^T, \sigma^2 I\right)$. Here, the origin is the mean, $\sigma^2 I$ is the covariance matrix, and $I$ is the unit matrix. Simulations that use this type of destination are henceforth referred to as Simulation 2.

The parameters were set to $\lambda = 1000.0$, $\sigma = 0.001$, $\alpha = 0.001$, and $l_{\max} = 10.0$. The initial position vector of the agent was set to $X = \left(0, 0\right)^T$. The simulation period was set to 200000 steps. However, the analyses presented below were performed using data from 100001 to 200000 steps.

# RESULTS

## Simulation results

Figures 3(a–c) and 4(a–c) depict the movement trajectories of the agents for Simulation 1 and 2,

respectively. It is important to note that the axis scales vary significantly in each figure.

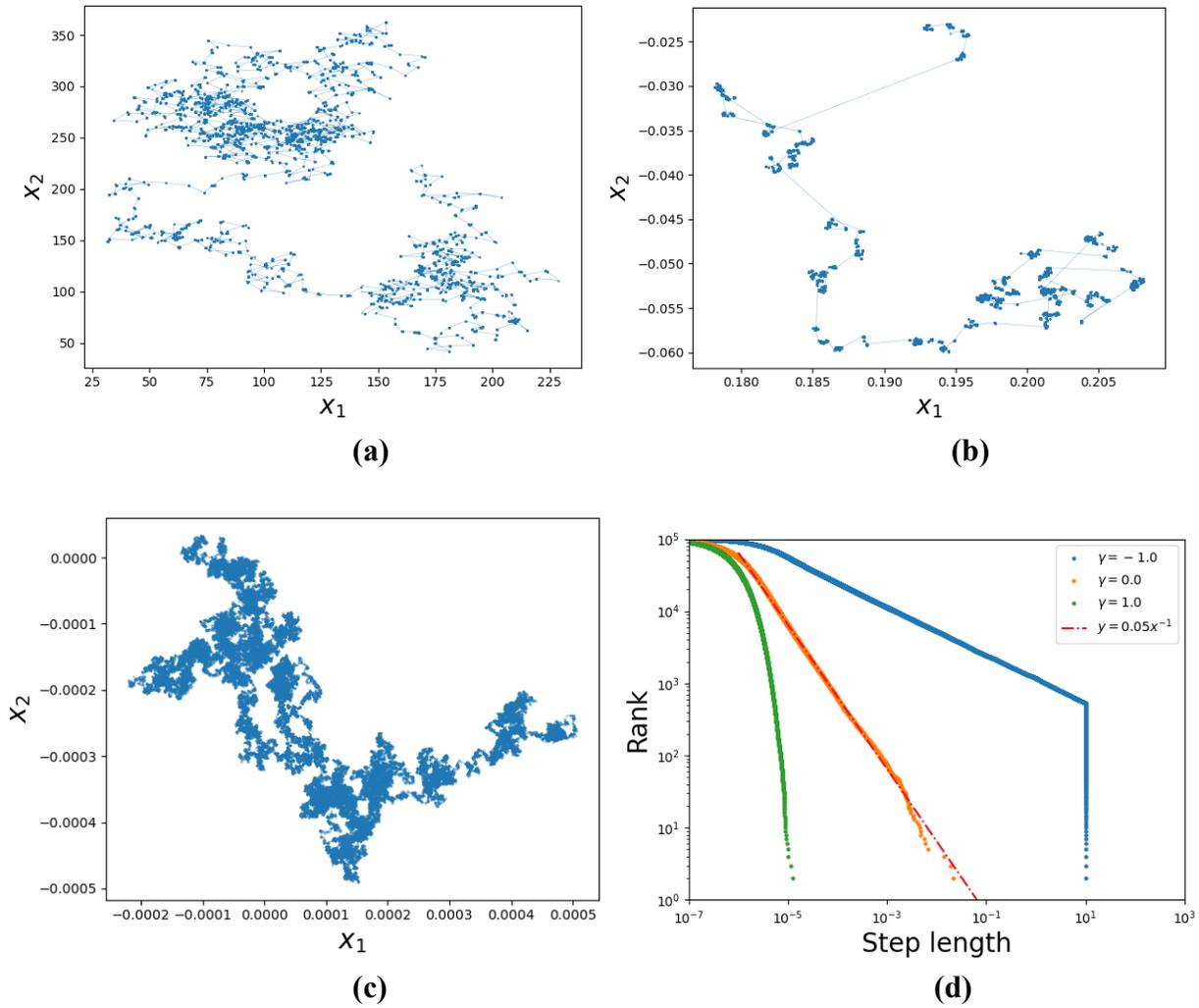

(a)

(b)

(c)

(d)

**Figure 3.** Movement trajectory of each agent in Simulation 1. **(a)** $\gamma = -1.0$, **(b)** $\gamma = 0.0$, **and (c)** $\gamma = 1.0$. **(d) Step length-rank plots for each agent in Simulation 1. Both axes are shown on a logarithmic scale.**

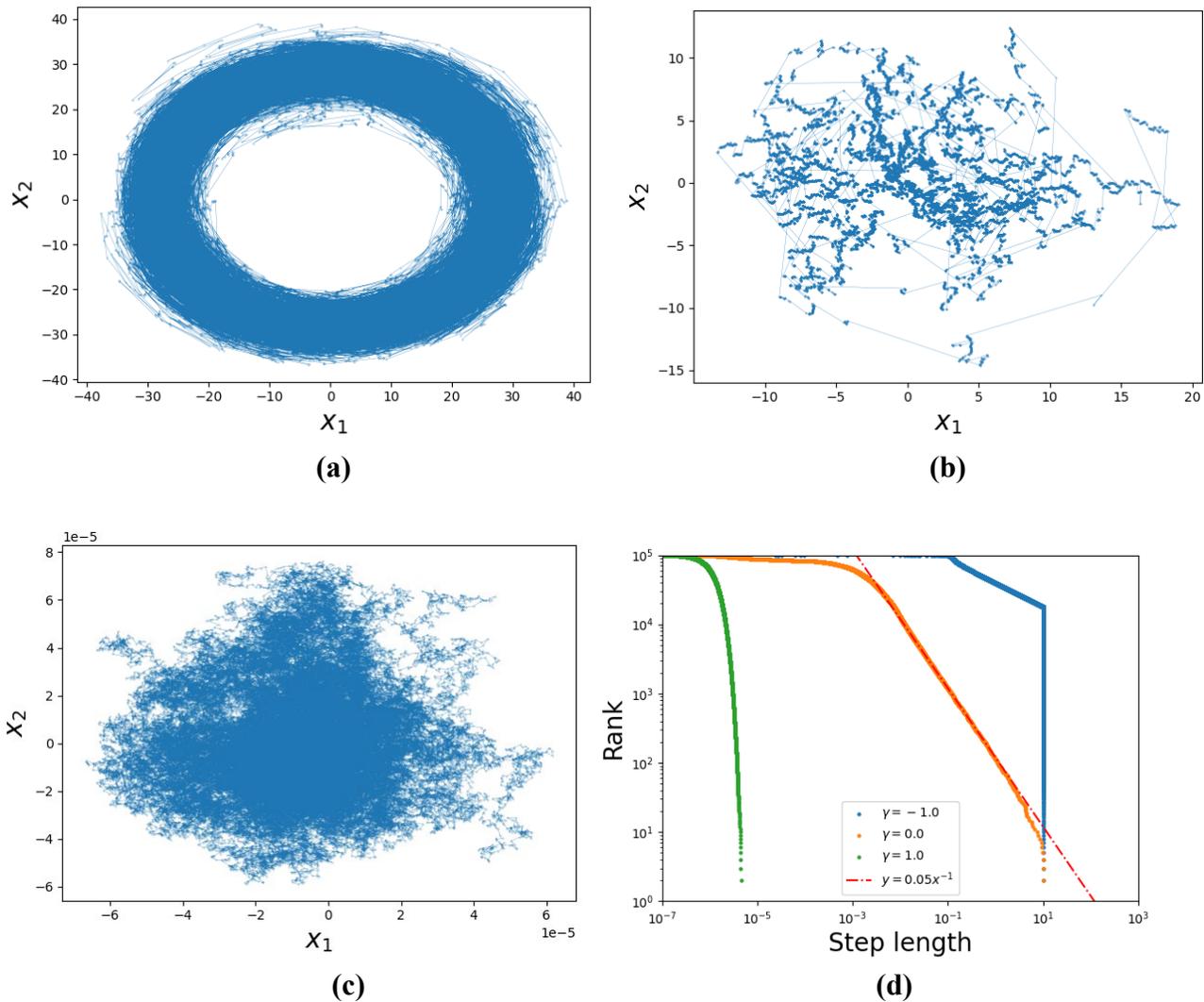

**(a)**

**(b)**

**(c)**

**(d)**

**Figure 4. Movement trajectory of each agent in Simulation 2. (a)** $\gamma = -1.0$**, (b)** $\gamma = 0.0$**, and (c)** $\gamma = 1.0$**. (d) Step length-rank plots for each agent in Simulation 2. Both axes are shown on a logarithmic scale.**

Figures 3(d) and 4(d) show the step length-rank plots of the three agents ($\gamma = -1.0$, $\gamma = 0.0$, $\gamma = 1.0$) for Simulations 1 and 2, respectively. These figures display step length on the horizontal axis and rank step lengths in descending order on the vertical axis. Note that no step length exceeds 10.0 because the maximum step length is limited to $l_{\max} = 10.0$. The rank of a step length indicates the number of step lengths that are greater than or equal to it, thereby representing a complementary cumulative distribution function (CCDF). These figures depict the power function with a dashed line.

The CCDF of the non-Min agent ($\gamma = 0.0$) can be approximated by the power function with exponent $-1$ in the long step length domain for both simulations. Thus, the step length distribution of the non-Min agent can be approximated by $P(l) \sim l^{-2}$ and the walking pattern can be considered a Cauchy walk.

The gait pattern of the agent with $\gamma = -1.0$ is a Lévy walk with power exponents less than two. As depicted in Figs. 3(a) and 4(a), the agent with $\gamma = -1.0$ searches a very large area. Interestingly, the agent does not approach the true destination—the origin—but orbits around the point in Simulation 2.

In both simulations, the slope of the power function becomes steeper as $\gamma$ increases. Since the step length of the Min agent ($\gamma = 1.0$) is theoretically $l = \alpha r$ per Eq. (6), $P(l)$ is the same as $P(r)$. Additionally, because the distance $r$ from the agent to the destination is sampled from the exponential distribution $P(r) = \lambda e^{-\lambda r}$ in Simulation 1, $P(l)$ is also an exponential distribution. Therefore, the walking pattern of the Min agent can be considered a Brownian walk.

Figure 5(a) shows the frequency with which the non-Min agent stays at each location on the $x_1 - x_2$ plane in Simulation 2. The frequency $N$ shown on the $z$-axis is the number of times the agent stays in each grid throughout the duration of Simulation 2, where the $x_1 - x_2$ plane is divided into grids of 0.02 per side. The frequency $N$ of each grid is the average of 1000 simulations. The figure shows a range of $[-10.0, 10.0]$ for both the $x_1$ and $x_2$ axes.

Let $r0 = \|X\| = \sqrt{\sum_{i=1}^{N}(x_i)^2}$ be defined as the distance between the agent and the origin, which is the fixed true destination in Simulation 2. Figure 5(b) shows the probability $P(r0 - \Delta r0 \le X < r0)$ that each agent stays in the region between $r0 - \Delta r0$ and $r0$ distance from the origin during Simulation 2. Herein, $\Delta r0 = 0.01$. The figure shows the results for the range of $0.01 \le r0 < 10.0$. The probability distribution was calculated from the frequency distribution of agents staying in each region after running the simulation 1000 times. The probability distribution of the non-Min agent ( $\gamma = 0.0$ ) can be approximated by a power distribution with exponent $-1$. This indicates that although the non-Min agent is more likely to stay in areas closer to the origin, it may extend its search area much further. This figure also shows that as $\gamma$ decreases, agents preferentially search farther from the origin, whereas as $\gamma$ increases, agents preferentially search closer to the origin. For $\gamma = -0.04$, the probability distribution is approximately flat, and the agent searches equally far and near.

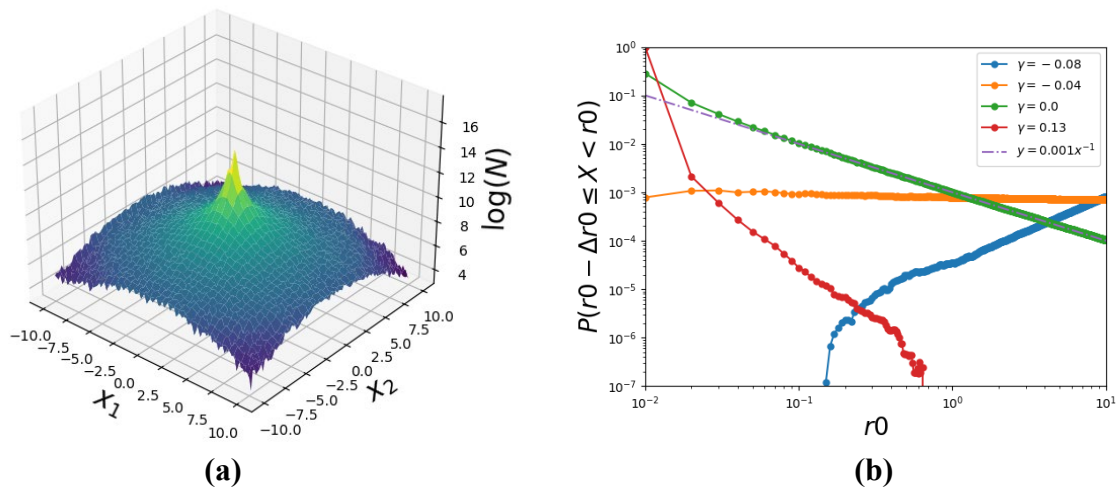

**(a)**  **(b)**

**Figure 5. Location where the agents stay during Simulation 2. (a) Frequency $N$ with which the non-Min agent ( $\gamma = 0.0$ ) stays at each location during Simulation 2. For clarity, the $Z$-axis is shown in $\log N$ . (b) Probability that the four agents ( $\gamma = -0.08$ , $\gamma = -0.04$ , $\gamma = 0.0$ , $\gamma = 0.13$ ) stay in the region between $r0 - \Delta r0$ and $r0$ distance from the origin during Simulation 2.**

Figures 6(a) and 6(b) show the time series of the distance $r0$ and the step-length for the non-Min agent ( $\gamma = 0.0$ ) in Simulation 2, respectively.

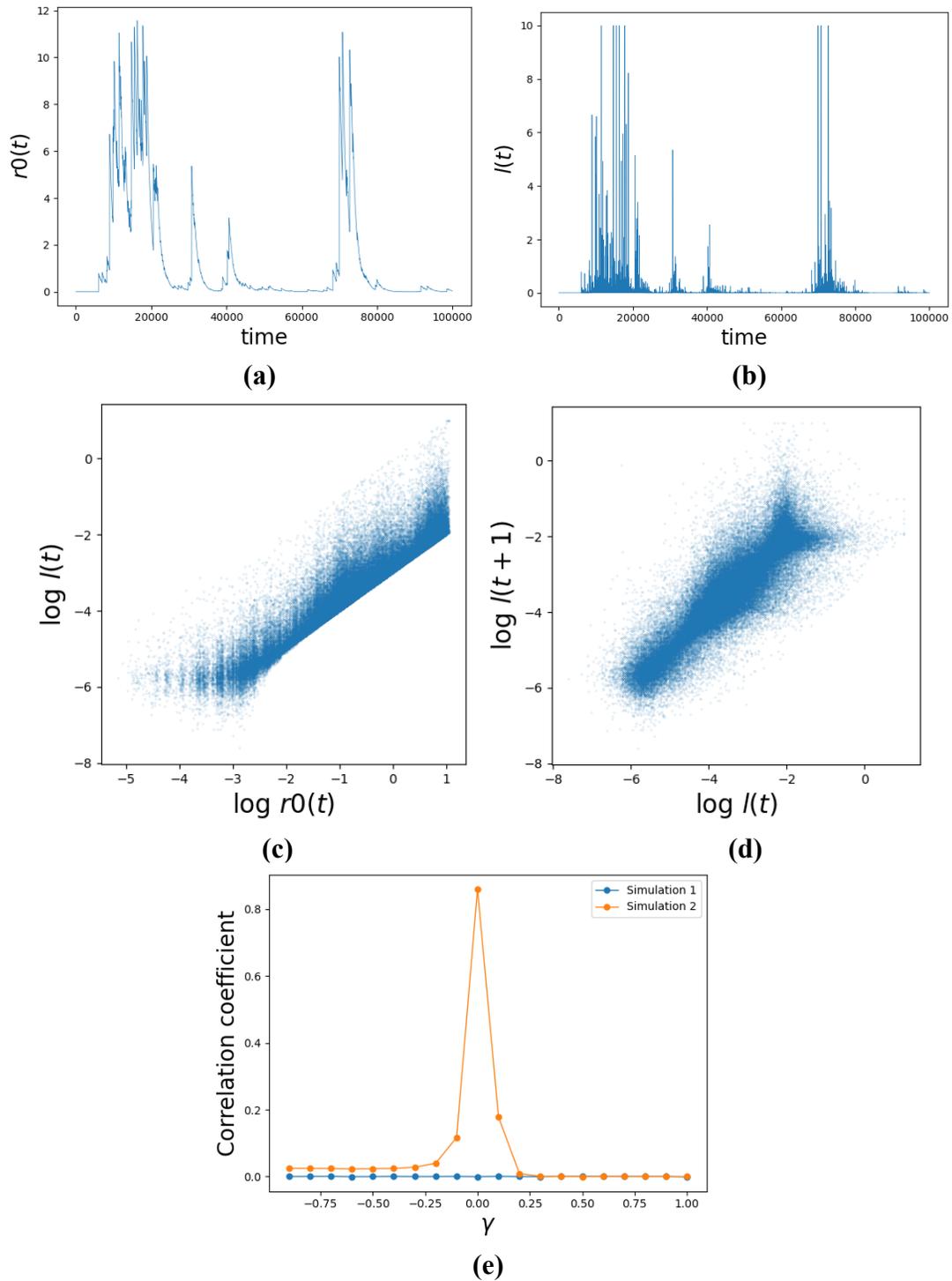

**(a)**

**(b)**

**(c)**

**(d)**

**(e)**

**Figure 6. Time series of $r0$ and $l$ for the non-Min agent ($\gamma = 0.0$) in Simulation 2. (a) Time series $r0(t)$ of $r0$. (b) Time series $l(t)$ of $l$. (c) Relationship between $r0(t)$ and $l(t)$. (d) Time dependence of $l(t)$. (e) Pearson correlation coefficients between $\log l(t)$ and $\log l(t+1)$ when $\gamma$ is varied from $-1$ to $1$ in $0.1$ increments.**

Figure 6(c) shows the relationship between the two time series $r0(t)$ and $l(t)$. The Pearson correlation coefficient between $\log r0(t)$ and $\log l(t)$ is 0.93. This indicates that the step length increases as the agent moves away from the origin. Figure 6(d) shows the time dependence of the step length, wherein the x-axis is $\log l(t)$ and the y-axis is $\log l(t+1)$. The Pearson correlation coefficient between them is 0.88. Figure 6(e) shows the Pearson correlation coefficients between $\log l(t)$ and $\log l(t+1)$ when $\gamma$ is varied from $-1$ to 1 in 0.1 increments. The results of Simulations 1 and 2 are shown together in the figure. The correlation coefficient for each $\gamma$ is the average of 100 simulations. Strong correlations are observed only in Simulation 2 near $\gamma = 0.0$.

## Cauchy walk analysis

In this section, we analyze why the Cauchy walk appears for the non-Min agent ($\gamma = 0.0$) in Simulation 1 and 2.

To simplify subsequent analysis, we denote $\beta_k = 1$, $\beta_j = 0 \, (j \neq k)$ for the non-Min agent. This indicates that the agent moves along the $k$-th axis and $\Delta x'_k \geq 0$, $\Delta x'_j = 0 \, (j \neq k)$. The direction of the $k$-axis is randomly determined each time, resulting in random movement direction.

If the angle between the $i$-th basis vector $e'_i$ and the difference vector $R'$ in the $N$-dimensional orthonormal system is $\theta_i$, then $R' = \left( r'_1, r'_2, \cdots, r'_N \right)^T = \left( r \cos \theta_1, \cdots, r \cos \theta_i, \cdots, r \cos \theta_N \right)^T$, since $\| R' \| = \| R \| = r$. Since $\beta_k = 1$ and $\beta_j = 0 \, (j \neq k)$, the step length can be written as follows.

$$l = \sqrt{\sum_{i=1}^{N} (\Delta x_i)^2}$$

$$\approx \sqrt{\sum_{i=1}^{N} \left( \alpha \beta_i \frac{r^2}{r'_i} \right)^2}$$

$$= \alpha \sqrt{\sum_{i=1}^{N} \left( \beta_i \frac{r}{\cos \theta_i} \right)^2}. \qquad (12)$$

$$= \alpha r \sqrt{\sum_{i=1}^{N} \left( \frac{\beta_i}{\cos \theta_i} \right)^2}$$

$$= \frac{\alpha r}{|\cos \theta_k|}$$

Consider $P(l \mid r) \propto \left| \frac{\partial \theta_k}{\partial l} \right| P(\theta_k)$, which is the distribution of $l$ when the current position $X$ is on the

hyper-sphere of radius $r$ centered at the destination $D$.

Since the step length of the non-Min agent is $l \approx \frac{\alpha r}{|\cos \theta_k|}$, it follows that $\left| \frac{\partial l}{\partial \theta_k} \right| \approx \alpha r \frac{|\sin \theta_k|}{\cos^2 \theta_k}$. Additionally,

$\cos^2 \theta_k = \frac{\alpha^2 r^2}{l^2}$ and $|\sin \theta_k| = \sqrt{1 - \cos^2 \theta_k} = \frac{\sqrt{l^2 - \alpha^2 r^2}}{l}$ also hold. Since $\theta_k$ is randomly determined, we can

put the probability of occurrence $P(\theta_k)$ of $\theta_k$ as a constant. Therefore, $P(l \mid r)$ can be written as

$$P(l \mid r) \propto \left| \frac{\partial \theta_k}{\partial l} \right| P(\theta_k)$$

$$= \frac{1}{\alpha r} \frac{\cos^2 \theta_k}{|\sin \theta_k|} P(\theta_k)$$

$$= \frac{1}{\alpha r} \frac{\alpha^2 r^2}{l^2} \frac{l}{\sqrt{l^2 - \alpha^2 r^2}} P(\theta_k). \qquad (13)$$

$$= \frac{\alpha r}{l \sqrt{l^2 - \alpha^2 r^2}} P(\theta_k)$$

$$\propto \frac{\alpha r}{l \sqrt{l^2 - \alpha^2 r^2}}$$

Finally, the distribution of $l$ can be written as follows.

$$P(l) \propto \int_r P(l \mid r) P(r) dr$$

$$\propto \int_r \frac{\alpha r}{l\sqrt{l^2 - \alpha^2 r^2}} P(r) dr \qquad (14)$$

As mentioned above, for the non-Min agent, $l \geq \alpha r$ is satisfied, and for $l$ such as $l \gg \alpha r$,

$l\sqrt{l^2 - \alpha^2 r^2} \approx l\sqrt{l^2} = l^2$ can be approximated. $P(l)$ can then be written as follows.

$$P(l) \propto \int_r \frac{\alpha r}{l\sqrt{l^2 - \alpha^2 r^2}} P(r) dr$$

$$\approx \int_r \frac{\alpha r}{l^2} P(r) dr$$

$$= \frac{\alpha}{l^2} \int_r r P(r) dr \qquad (15)$$

$$\propto \frac{1}{l^2}$$

Thus, we see that $P(l)$ of the non-Min agent can approximate the Cauchy distribution regardless of $P(r)$

with respect to $l$ such that $l \gg \alpha r$. Since $l = \alpha r$ for the Min agent, $P(l)$ has the same distribution as $P(r)$.

# DISCUSSION

First, we constructed a walking model that continuously generates a Brownian walk to a Lévy walk by varying parameter $\gamma$. The simulation results show that the Cauchy walk appears when $\gamma = 0.0$; i.e., when the parameters $\beta_i$ that determine the allocation of the amount of movement to each axis are set randomly. Conversely, for the agent's gait to be anything other than a Cauchy walk, some adjustment must be made to the parameters.

Next, in Simulation 2, with the true destination fixed at the origin, a time correlation was found in the step-length time series of the agents around $\gamma = 0.0$. For the non-Min agents with $\gamma = 0.0$, the step length tends to increase as the distance from the origin increases, as shown in Fig. 6(c). The change rate is small ($\alpha = 0.001$), and once an agent is far away from the origin, it will gradually move closer to the origin. This explains the time correlation observed for the non-Min agent. Regarding the Min agent ($\gamma = 1.0$), it fluctuates around the origin due to some identification error, as illustrated in Figure 4(c). Consequently, there is no temporal correlation in the step-length time series for the Min agents. Previous reporting indicates that a temporal correlation can be observed in the step-length time series for human movement, and that the closer the gait pattern is to a Lévy walk, the stronger the temporal correlation.[25] Our simulation results are consistent with these findings.

The proposed model is abstract, and future work will be necessary to understand the biological implications. In this model, we can generate various migratory behaviors, including Brownian walks and Lévy walks, by controlling parameter $\gamma$. Parameter $\gamma$ represents the attraction force of the destination for agents, and the larger $\gamma$, the stronger the force. When the destination is fixed as in Simulation 2, the destination with large $\gamma$ may be considered a nest or point where the agent has acquired more food previously.

If there is an abundance of food in the vicinity of the destination and the agent considers the area attractive, it should increase $\gamma$ and focus the search on that area, using Brownian walk. Conversely, if the agent cannot find food in that area, it should reduce $\gamma$, abandon the area, and explore further afield. As shown in Fig. 5(b), the non-Min agent with $\gamma = 0.0$ lies exactly in the middle of the two and searches the region with a probability inversely proportional to the distance from the destination. Therefore, the Cauchy walk can be seen in the non-Min agent presumably when it is unable to judge whether the area around the destination is attractive or unattractive.

In areas where food is scarce, marine predators tend to adopt the Lévy walk, whereas in regions with abundant food, they shift to the Brownian walk[4]. Future work will include creating an agent model that adjusts $\gamma$ as the environment changes.

For the simulations in this study, two types of destinations were set up. If we assume the destination to be the location of another agent, agents with positive $\gamma$ will attempt to approach each other, or agents with negative $\gamma$ will attempt to avoid each other, potentially allowing for the construction of a flocking model. Additionally, this model, being a walk model generating a Cauchy walk in multi-dimensional space, could potentially be used to search the parameter space of an optimization problem.[26] These potential applications will be explored in future work.

# Acknowledgments

This work was supported by JSPS KAKENHI [grant number JP21K12009].

# Author contributions

**Shuji Shinohara:** Conceptualization, formal analysis, methodology, software, writing, original draft preparation, and funding acquisition. **Daiki Morita:** Software, reviewing, and editing. **Nobuhito Manome:** Writing, reviewing, and editing. **Hayato Hirai:** Software, review, and editing. **Ryosuke Kuribayashi:** Software, review, and editing. **Toru Moriyama:** Writing, reviewing, and editing. **Hiroshi Okamoto:** Writing, reviewing, and editing. **Yoshihiro Nakajima:** Writing, reviewing, and editing. **Pegio-Yukio Gunji:** Writing, reviewing, editing, and supervision. **Ung-il Chung:** Writing, review, editing, supervision, and project administration.